\documentclass[12pt]{article}
\usepackage{graphicx}
\usepackage{lscape}
\usepackage{epsfig}
\usepackage{citesort}
\usepackage{amssymb}
\usepackage{amsmath}
\usepackage{multirow}
\newcommand{\be}{\begin{equation}}
\newcommand{\ee}{\end{equation}}
\newcommand{\bea}{\begin{eqnarray}}
\newcommand{\eea}{\end{eqnarray}}

\def\R1{\varepsilon_1}
\def\E8{\varepsilon_8}

\def\ga{\gamma}

\def\s1{\hat s}
\def\ds{\displaystyle}

\newcommand{\bd}{\begin{displaymath}}
\newcommand{\ed}{\end{displaymath}}

\newcommand{\f}{\frac}

\def\R1{\varepsilon_1}
\def\E8{\varepsilon_8}

\def\ga{\gamma}

\def\ds{\displaystyle}
\def\beq{\begin{equation}}
\def\eeq{\end{equation}}
\def\bea{\begin{eqnarray}}
\def\eea{\end{eqnarray}}
\def\beeq{\begin{eqnarray}}
\def\eeeq{\end{eqnarray}}

\def\vel{\left|}
\def\ver{\right|}
\def\nnb{\nonumber}
\def\ga{\left(}
\def\dr{\right)}

\def\lla{\left<}
\def\rra{\right>}
\def\rar{\rightarrow}
\def\nnb{\nonumber}

\def\ba{\begin{array}}
\def\ea{\end{array}}

\def\xis0{{\Xi^{*0}}}

\def\g5{\gamma_5}

\def\es{\!\!\! &=& \!\!\!}

\def\ek{&-& \!\!\!}

\newcommand{\al}{\alpha_s}

\begin{document}

\title{
         {\Large
                 {\bf
                     Double-Lepton Polarization
Asymmetries and Branching Ratio  of the $B\rar \gamma l^+ l^- $
transition in Universal Extra Dimension
                 }
         }
      }
\author{\vspace{1cm}\\
{\small  K. Azizi$^1$ \thanks {e-mail: kazizi@dogus.edu.tr}\,\,, N. K. Pak$^2$ \thanks
{e-mail:pak@metu.edu.tr}\,\,, B. B.  \c{S}irvanl{\i}$^3$\thanks
{e-mail:bbelma@gazi.edu.tr}}  \\
{\small $^1$ Department of Physics, Do\u gu\c s University,
Ac{\i}badem-Kad{\i}k\"oy, 34722 Istanbul, Turkey}\\
{\small $^2$ Department of Physics, Middle East Technical University,
06800 Ankara, Turkey}\\
{\small $^3$ Department of Physics,  Gazi University,  Teknikokullar,  06100 Ankara, Turkey}}
\date{}

\begin{titlepage}
\maketitle
\thispagestyle{empty}

\begin{abstract}
We study  the radiative dileptonic $B \rar \gamma l^+ l^- $ transition  in the presence of a universal
extra dimension in the
Applequist-Cheng-Dobrescu  model.  In particular, using the corresponding form factors calculated via light cone QCD sum rules, we analyze the  branching ratio and  double lepton polarization
asymmetries related to this channel  and
compare the  results with the predictions of the standard
model. We show  how the results deviate from predictions of the standard model  at lower values of the compactification factor ($1/R $)
 of extra dimension. 

\end{abstract}

~~~PACS number(s): 12.60-i, 13.20.-v, 13.20.He
\end{titlepage}

%%%
\section{Introduction \label{s1}}
Although the standard model (SM) of particle physics is in perfect agreement with all confirmed collider data,  there are some problems that can not be addressed by the SM. Some of these
 problems are  matter-antimatter asymmetry, number of generations, unification of the fundamental interactions, etc. Hence, we need more fundamental theories beyond the SM (BSM)
 such that at low energies those theories reduce to the SM. One of the most interesting candidates as a BSM theory is extra dimension (ED) \cite{Antoniadis1,Antoniadis2,Arkani1,Arkani2,Randall1,Randall2}. 
A kind of ED which permits both gauge bosons and fermions as SM fields to spread in ED's is labeled as universal extra dimension (UED). The  simplest case of the UED is the Applequist-Cheng-Dobrescu (ACD)
model \cite{ACD} which contains only one UED compactified in a circle of radius $R$.

We have no experimental evidence for  the new physics effects such as ED's so far, but we expect  that the LHC will open new horizons in this respect. 
 There are two alternative ways to search for ED's. In  direct search, we look for  Kaluza-Klein (KK) excitations directly by increasing  the center of mass energy of colling particles. In indirect search, we look for the
contributions of the KK particles to the hadronic decay channels. The flavor changing neutral current (FCNC) transitions induced by  loop level quark transitions are considered as good tools for studying the
KK effects.

The ACD model has been previously applied to many rare semileptonic  decay channels \cite{R7624,wangying,R7601,wangying,sirvanli,kank1,bashiry,carlucci,aliev,ahmet,nihan}. In the present work, we apply this model 
to analyze the  branching ratio and  double lepton polarization asymmetries defining the radiative dileptonic $B \rar \gamma l^+ l^- $ transition. The advantage of such decay channel compared to the pure leptonic
helicity suppressed   $B\rar  \mu^+ \mu^- $ and $B \rar  e^+ e^- $ channels is that due to the emission of the  photon in addition to the lepton pair,  we have no helicity suppression
 here and we expect larger branching ratio \cite{Aliev1,Aliev2}. The upper experimental limits, $Br(B \rar  \mu^+ \mu^-)<1.5\times10^{-8}$, $Br(B \rar \gamma \mu^+ \mu^-)<1.6\times10^{-7}$, 
$Br(B \rar  e^+ e^-)<8.3\times10^{-8}$, $Br(B \rar \gamma e^+ e^-)<1.2\times10^{-7}$ \cite{Nakamura} verify our expectations in this respect.  The considered decay channel proceeds via FCNC transition of $b \rar d l^+ l^-$ at quark level and as we previously mentioned the KK particles can 
contribute to such channels. To evaluate the branching ratio and various double lepton polarization asymmetries, we will use the  form factors entering the effective Hamiltonian calculated via 
light cone QCD sum rules \cite{Aliev1,Eilam2,Alievsav,Buchalla}.

The layout of the paper is as follows. The introduction is followed by section 2 which encompasses the theoretical background of the  decay channel under consideration and 
the associated effective Hamiltonian, a brief review on the ACD model, transition matrix elements defining the radiative dileptonic
 $B \rar \gamma l^+ l^- $ decay channel and explicit expressions for the associated observables (differential decay rate and double lepton polarization asymmetries) in the UED model. In section 3, 
using the fit parametrization of form factors as the main ingredients as well as other input parameters, we numerically analyze the physical observables both in the UED and the SM models and discuss how the results obtained from
 the UED model deviate from those of the SM.

\section{The   radiative dileptonic $B\rar \gamma l^+ l^- $  transition in the ACD model\label{s2}} 
As we mentioned in the previous section,  the $B\rar \gamma l^+ l^- $
transition proceeds via the FCNC transition of the $b \rar d l^+ l^- $  at the quark level. The most important contribution to the $B\rar \gamma l^+ l^- $ comes from the pure leptonic 
 $B\rar  l^+ l^- $ transition. The latter proceeds via the box and $Z$-photon mediated penguin diagrams (see for instance \cite{Aliev1,Aliev2}). By attaching the photon to any external and internal charged lines,
we will obtain the transition matrix elements for the $B\rar \gamma l^+ l^- $ decay. In the SM, the effective Hamiltonian responsible for  $b\rar q \ell^+ \ell^-$ transition can be written as
 \bea \label{e6801} 
{\cal H}_{eff} \es
\frac{\alpha G_F}{\sqrt{2} \pi} V_{tb} V_{td}^*
\Bigg{[} C_9^{eff} (\bar d \gamma_\mu P_L b) \bar l \gamma_\mu l +
C_{10}\bar d \gamma_\mu P_L b \bar l \gamma_\mu \gamma_5 l  - 2 \frac{C_7^{eff}}{q^2} \bar d i \sigma_{\mu \nu} q_\nu ( m_b P_R + m_d P_L ) b
\bar l \gamma_\mu l 
\Bigg{]},\nnb\\ \eea 
where $P_{R(L)} =\frac{ 1 + (-) \gamma_5 }{2}$, and $q^2$  is the transferred  momentum squared. The $ C_7^{eff}$, $ C_9^{eff}$ and $ C_{10}$ are Wilson coefficients which are the source of difference between the SM
 and UED models. In the UED, the form of Hamiltonian remains unchanged; however, the Wilson coefficients are modified \cite{Buras:2002ej,R7623,R7626,R7627,R762777} 
as a result of  interactions of the KK particles  with each other as well as with the usual SM particles. In this model, each Wilson coefficient is written in
 terms of the ordinary SM part and an extra part coming
 from the aforementioned interactions. Hence, 
\begin{equation}
 F(x_t,1/R)=F_0(x_t)+\sum_{n=1}^{\infty}F_n(x_t,x_n),
\label{functions}
\end{equation}
 where $F_0 (x_t )$ is the SM part and  $x_{t}=m_{t}^{2}/M_{W}^{2}$ with $m_{t}$ and  $M_{W}$ being   masses of the top quark and the $W$ boson, respectively. The second part is defined in terms of the
compactification factor $1/R$  via
\begin{equation}
 x_n=m_n^2/m_W^2,~~~~~~\mbox{with}~~~~~~~~m_n=n/ R,
\end{equation}
  where  $m_n$ is   mass of the KK particles and $n=0$ corresponds to the ordinary SM  particles. Here we should also mention that the  KK sums appearing in all Wilson coefficients converge and give finite results.

Now, we proceed to present the explicit expressions of the  Wilson coefficients entering the low energy effective Hamiltonian obtained by a renormalization 
group evolution from the electroweak scale down to the
 $m_b$ scale. In the leading log approximation, the coefficient  $C_7^{eff}(1/R)$   is written as  \cite{Buras:2002ej,R7623,R7626,R7627,R762777}:
  \bea
\label{e7603} C_7^{eff}(\mu_b, 1/R) \es \eta^{\frac{16}{23}}
C_7(\mu_W, 1/R)+ \frac{8}{3} \left( \eta^{\frac{14}{23}}
-\eta^{\frac{16}{23}} \right) C_8(\mu_W, 1/R)+C_2 (\mu_W)
\sum_{i=1}^8 h_i \eta^{a_i}~, \nnb\\ \eea 
where
 \bea \eta \es
\frac{\alpha_s(\mu_W)} {\alpha_s(\mu_b)}~,\eea
and
 \bea
\alpha_s(x)=\frac{\alpha_s(m_Z)}{1-\beta_0\frac{\alpha_s(m_Z)}{2\pi}\ln(\frac{m_Z}{x})},\eea
with $\alpha_s(m_Z)=0.118$ and $\beta_0=\frac{23}{3}$. The coefficients $a_i$ and $h_i$, with $i$ running from $1$ to $8$, are also  given as \cite{R7627,R762777}:
   \be
   \label{klar}
\begin{array}{rrrrrrrrrl}
a_i = (\!\! & \f{14}{23}, & \f{16}{23}, & \f{6}{23}, & -
\f{12}{23}, &
0.4086, & -0.4230, & -0.8994, & 0.1456 & \!\!)  \vspace{0.1cm}, \\
h_i = (\!\! & 2.2996, & - 1.0880, & - \f{3}{7}, & - \f{1}{14}, &
-0.6494, & -0.0380, & -0.0186, & -0.0057 & \!\!).
\end{array}
\ee 
The functions
\bea
 C_2(\mu_W)=1~,~~ C_7(\mu_W, 1/R)=-\frac{1}{2}
D^\prime(x_t,1/R)~,~~ C_8(\mu_W, 1/R)=-\frac{1}{2}
E^\prime(x_t,1/R)~ . \eea
Also, the $1/R$ -dependent functions  $D^\prime (x_t,1/R)$  and $E^\prime (x_t,1/R)$ are defined as
\bea
D^\prime (x_t,1/R)=D^\prime_0(x_t)+\sum_{n=1}^{\infty}D^\prime_n(x_t,x_n),~~~~~E^\prime (x_t,1/R)=E^\prime_0(x_t)+\sum_{n=1}^{\infty}E^\prime_n(x_t,x_n)~, \eea
where the SM parts are given as
 \bea \label{e7604} D^\prime_0(x_t) \es - \frac{(8
x_t^3+5 x_t^2-7 x_t)}{12 (1-x_t)^3}
+ \frac{x_t^2(2-3 x_t)}{2(1-x_t)^4}\ln x_t~, \\ \nnb \\
\label{e7605} E^\prime_0(x_t) \es - \frac{x_t(x_t^2-5 x_t-2)}{4
(1-x_t)^3} +
\frac{3 x_t^2}{2 (1-x_t)^4}\ln x_t~, 
\eea
 and the parts coming from the new interactions can be written in the forms
 \bea \label{e7608}
&&\sum_{n=1}^{\infty}D^\prime_n(x_t,x_n) \nnb \\ \es
\frac{x_t[37 - x_t(44+17 x_t)]}{72 (x_t-1)^3} 
+ \frac{\pi m_W R}{12} \Bigg[ \int_0^1 dy \, (2 y^{1/2}+7
y^{3/2}+3 y^{5/2}) \, \coth (\pi m_WR \sqrt{y}) \nnb \\
\ek \frac{x_t (2-3 x_t) (1+3 x_t)}{(x_t-1)^4}J(R,-1/2)
- \frac{1}{(x_t-1)^4} \Big\{ x_t(1+3 x_t)+(2-3 x_t)
[1-(10-x_t)x_t] \Big\} \nnb \\
&\times& J(R, 1/2)- \frac{1}{(x_t-1)^4} [ (2-3 x_t)(3+x_t) + 1 - (10-x_t) x_t]
J(R, 3/2)- \frac{(3+x_t)}{(x_t-1)^4} J(R,5/2)\Bigg]~, \nnb \\
\eea
and
\bea
\label{e7609} &&\sum_{n=1}^{\infty}E^\prime_n(x_t,x_n)\nnb \\\es
\frac{x_t[17+(8-x_t)x_t]}
{24 (x_t-1)^3} + \frac{\pi m_W R}{4} \Bigg[\int_0^1 dy \, (y^{1/2}+
2 y^{3/2}-3 y^{5/2}) \, \coth (\pi m_WR \sqrt{y}) \nnb \\
\ek {x_t(1+3 x_t) \over (x_t-1)^4}J(R,-1/2)+ \frac{1}{(x_t-1)^4} [ x_t(1+3 x_t) - 1 + (10-x_t)x_t] J(R, 1/2)\nnb \\
\ek \frac{1}{(x_t-1)^4} [(3+x_t)-1+(10-x_t)x_t) ]J(R, 3/2)+{(3+x_t) \over  (x_t-1)^4} J(R,5/2)\Bigg]~. \eea 
Here, \bea
\label{e76010} J(R,\alpha)=\int_0^1 dy \, y^\alpha \left[ \coth
(\pi m_W R \sqrt{y})-x_t^{1+\alpha} \coth(\pi m_t R \sqrt{y})
\right]~. \eea

The next Wilson coefficient is $C_9^{eff}$.  In the leading log approximation and at $\mu_b$ scale it is given as \cite{R7627,R762777}:
 \bea \label{C9eff}
C_9^{eff}(\mu_b,\hat{s}',1/R) & = & C_9^{NDR}(1/R)\eta(\hat s') + h(z, \hat s')\left( 3
C_1 + C_2 + 3 C_3 + C_4 + 3
C_5 + C_6 \right) \nonumber \\
& & - \f{1}{2} h(1, \hat s') \left( 4 C_3 + 4 C_4 + 3
C_5 + C_6 \right) \nonumber \\
& & - \f{1}{2} h(0, \hat s') \left( C_3 + 3 C_4 \right)
+ \f{2}{9} \left( 3 C_3 + C_4 + 3 C_5 +
C_6 \right), \eea
where, $\hat{s}'=\frac{q^2}{m_b^2}$ with the physical region $4m_l^2\leq q^2\leq m_{B}^2$. The function $C_9^{NDR}(1/R)$ in  the naive dimensional regularization (NDR) scheme is defined as
\bea \label{C9tilde}C_9^{NDR}(1/R) & = & P_0^{NDR} +
\f{Y(x_t)}{\sin^2\theta_W} -4 Z(x_t) + P_E E(x_t). \eea 
 Here we should underline that, due to  smallness  of $P_E$, we can neglect the contribution of last term in Eq. (\ref{C9tilde}). The 
 constant $P_0^{NDR}=2.60 \pm 0.25$ \cite{R7627,R762777}, and remaining two functions
$Y(x_t,1/R)$ and $Z(x_t,1/R)$  have the following expressions:
\bea
\label{e7612}
Y(x_t,1/R) \es Y_0(x_t)+\sum_{n=1}^\infty C_n(x_t,x_n)~, 
\eea where, \bea
\label{e7613} Y_0(x_t) \es \frac{x_t}{8} \left[ \frac{x_t -4}{x_t
-1}+\frac{3 x_t}{(x_t-1)^2} \ln x_t \right]~, \eea  and,  \bea
\label{e7615} \sum_{n=1}^\infty C_n(x_t,x_n) = \frac{x_t(7-x_t)}{
16 (x_t-1)} - \frac{\pi m_W R x_t}{16 (x_t-1)^2}
\left[3(1+x_t)J(R,-1/2)+(x_t-7)J(R,1/2) \right]~.\nnb\\ \eea 
also
\bea \label{e7612}
 Z(x_t,1/R) \es Z_0(x_t)+\sum_{n=1}^\infty
C_n(x_t,x_n)~, \eea with \bea Z_0(x_t) \es \frac{18 x_t^4-163
x_t^3+259 x_t^2 -108 x_t}{144 (x_t-1)^3} \nnb +\left[\frac{32
x_t^4-38 x_t^3-15 x_t^2+18 x_t}{72
(x_t-1)^4} - \frac{1}{9}\right] \ln x_t .\\ \nnb \\
\eea 
To complete the presentation of the coefficient $C_9^{eff}$ in Eq. (\ref{C9eff}), we define
\bea \eta(\hat s') & = & 1 +
\f{\al(\mu_b)}{\pi}\, \omega(\hat s'), \eea
where,
 \bea \label{omega}
\omega(\hat s') & = & - \f{2}{9} \pi^2 - \f{4}{3}\mbox{Li}_2(\hat s') -
\f{2}{3}
(\ln \hat s') \ln(1-\hat s') - \f{5+4\hat s'}{3(1+2\hat s')}\ln(1-\hat s') - \nonumber \\
& &  \f{2 \hat s' (1+\hat s') (1-2\hat s')}{3(1-\hat s')^2 (1+2\hat s')} \ln \hat s' + \f{5+9\hat s'-6\hat s'^2}{6
(1-\hat s') (1+2\hat s')}. \eea 
The coefficients  $C_j~(j=1,...6) $ are given as
 \bea \label{coeffs} C_j=\sum_{i=1}^8 k_{ji}
\eta^{a_i} \qquad (j=1,...6) \vspace{0.2cm} \eea and the constants $k_{ji}$
 have the values
\be\frac{}{}
   \label{klar}
\begin{array}{rrrrrrrrrl}
k_{1i} = (\!\! & 0, & 0, & \f{1}{2}, & - \f{1}{2}, &
0, & 0, & 0, & 0 & \!\!),  \vspace{0.1cm} \\
k_{2i} = (\!\! & 0, & 0, & \f{1}{2}, &  \f{1}{2}, &
0, & 0, & 0, & 0 & \!\!),  \vspace{0.1cm} \\
k_{3i} = (\!\! & 0, & 0, & - \f{1}{14}, &  \f{1}{6}, &
0.0510, & - 0.1403, & - 0.0113, & 0.0054 & \!\!),  \vspace{0.1cm} \\
k_{4i} = (\!\! & 0, & 0, & - \f{1}{14}, &  - \f{1}{6}, &
0.0984, & 0.1214, & 0.0156, & 0.0026 & \!\!),  \vspace{0.1cm} \\
k_{5i} = (\!\! & 0, & 0, & 0, &  0, &
- 0.0397, & 0.0117, & - 0.0025, & 0.0304 & \!\!) , \vspace{0.1cm} \\
k_{6i} = (\!\! & 0, & 0, & 0, &  0, &
0.0335, & 0.0239, & - 0.0462, & -0.0112 & \!\!).  \vspace{0.1cm} \\
\end{array}
\ee
Finally, we should define the other functions in Eq. (\ref{C9eff}):
 \bea \label{phasespace} h(y,
\hat s') & = & -\f{8}{9}\ln\f{m_b}{\mu_b} - \f{8}{9}\ln y +
\f{8}{27} + \f{4}{9} x \\
& & - \f{2}{9} (2+x) |1-x|^{1/2} \left\{
\begin{array}{ll}
\left( \ln\left| \f{\sqrt{1-x} + 1}{\sqrt{1-x} - 1}\right| - i\pi
\right), &
\mbox{for } x \equiv \f{4z^2}{\hat s'} < 1 \nonumber \\
2 \arctan \f{1}{\sqrt{x-1}}, & \mbox{for } x \equiv \f {4z^2}{\hat
s'} > 1,
\end{array}
\right. \\
\eea 
where   $y=1$ or $y=z=\frac{m_c}{m_b}$ and,
\bea h(0, \hat s') & = & \f{8}{27}
-\f{8}{9} \ln\f{m_b}{\mu_b} - \f{4}{9} \ln \hat s' + \f{4}{9}
i\pi.\eea

The  Wilson coefficient $C_{10}$ is scale-independent and is given  as:
 \bea
\label{e7616} C_{10}(1/R)= - \frac{Y(x_t,1/R)}{\sin^2 \theta_W}~, \eea
where, $\sin^2\theta_W= 0.23$.

Once the  Wilson coefficients in UED model are specified explicitly, we proceed to obtain the amplitude for the  decay channel under consideration which is obtained by sandwiching the effective Hamiltonian between 
the final photon and the 
initial $B$ meson state. As  previously noted,  the diagrams defining the  $B\rar \gamma l^+ l^- $ transition are obtained attaching the photon to any external and internal charged lines. 
Hence, we have 
three kinds of contributions: 1)  the photon is emitted from the initial quark lines, 2)  the photon is radiated from the final charged lepton lines
 and 3)  the photon is attached to any charged internal line. When photon is attached to the initial quark lines (structure dependent part), the  $B\rar \gamma l^+ l^- $ transition 
 is described by three Wilson coefficients  $ C_7^{eff}$, $ C_9^{eff}$ and $ C_{10}$ and we deal with the long
distance effects. Therefore, the amplitude is written as 
 \bea
M_1=\langle \gamma(k)|{\cal H}_{eff}|B(p)\rangle
 \eea
where k is the momentum of the photon and the $p=k+q$ is the initial momentum. 
To obtain the  amplitude $M_1$, we need to   define the matrix elements
 \bea \label{e6804} \lla \gamma(k) \vel \bar d
\gamma_\mu (1 - \gamma_5) b \ver B(p) \rra \es \frac{e}{m_B^2}
\Big\{ \epsilon_{\mu\nu\lambda\sigma} \varepsilon^{\ast\nu}
q^\lambda k^\sigma g(q^2) + i\, \Big[ \varepsilon^{\ast}_\mu (k
q) - (\varepsilon^\ast q) k_\mu \Big] f(q^2) \Big\}~,\nnb\\\eea 
and
 \bea \label{e6808} \lla \gamma(k) \vel
\bar d i \sigma_{\mu\nu} q^\nu (1+\gamma_5) b \ver B (p)\rra \es
\frac{e}{m_B^2} \Big\{ \epsilon_{\mu\nu\lambda\sigma} \,
\varepsilon^{\nu\ast} q^\lambda k^\sigma g_1(q^2) +
i\,\Big[\varepsilon_\mu^\ast (q k) - (\varepsilon^\ast q) k_\mu
\Big] f_1(q^2) \Big\}~, \nnb\\\eea 
where $\varepsilon^\ast_\mu$ is the four vector polarization  of the photon, and $g(q^2)$, $f(q^2)$, $g_1(q^2)$ and $f_1(q^2)$ are the transition form factors.

When photon is radiated from the final charged leptons (Bremsstrahlung part) the corresponding amplitude is called $M_2$. From the helicity arguments it follows that 
the amplitude $M_2$ should be proportional to the lepton mass $m_l$; for the cases of the $l=e,~\mu$ we can safely ignore from such contributions. For $\tau$ lepton case, this amplitude is calculated in \cite{Aliev2}. Finally, when the photon is attached to any charged internal line, the 
amplitude of such contributions ($M_3$) is proportional to $\frac{m_b^2}{m_W^2}$; so these contributions for all leptons are strongly suppressed and we can safely ignore those contributions (see for instance \cite{Aliev1,Aliev2}).

Now we proceed to present the $1/R$-dependent physical observables defining the radiative dileptonic $B \rar \gamma l^+ l^- $ transition. Considering the aforementioned contributions, the differential decay rate
for the $l=e$ or $\mu$ case as a function of the compactification factor is obtained as \cite{Aliev1}:
\bea \label{kab} \frac{d \Gamma}{d \hat{s}}(\hat{s},1/R) &=& \frac{\alpha^3
G_{F}^2}{768 \pi^5} \left| V_{tb} V_{td}^* \right|^2 m_B^5 \hat{s}
( 1-\hat{s})^3
\sqrt{1 - 4 \frac{\hat{m_{l}}^2}{\hat{s}}} \times \nnb \\
&&\times \Bigg{\{} \frac{1}{m_B^2} \left[ \left| A \right|^2 +
\left| B \right|^2 \right] +\frac{1}{m_B^2} \left| C_{10}(1/R)
\right|^2 \left[ f^2(q^2) + g^2(q^2) \right] \Bigg{\}}~, \eea

where $\hat{s}=\frac{q^2}{m_B^2}$, $\hat{m_l}=\frac{m_l}{m_B}$,
\bea
A = A (\hat{s},1/R)&=&  C_9^{eff}(\hat{s},1/R) g(q^2) - 2 \, C_7^{eff}(1/R) \frac{m_b}{\hat{s}m_B^2} g_1(q^2)~,\nnb \\ \mbox{and} \nnb \\
B = B (\hat{s},1/R)&=& C_9^{eff}(\hat{s},1/R) f(q^2) - 2 \,
C_7^{eff}(1/R) \frac{m_b}{\hat{s}m_B^2} f_1(q^2)~. \nnb \eea

In the case of $\tau$, the differential decay width is obtained as \cite{Aliev2}:
\bea
\label{kabciv}&& \frac{d \Gamma}{d \hat{s}}(\hat{s},1/R)=\nnb \\ && \vel \frac{\alpha G_F}{2 \sqrt{2} \, \pi} V_{tb} V_{td}^* \ver^2 \,
\frac{\alpha}{\ga 2 \, \pi \dr^3}\, m_B^5 \pi  \Bigg{\{} \frac{1}{12} \, \int_\delta^{1-4 r} x^3 \,dx \,
\sqrt{1-\frac{4 r}{1 - x}} \, m_B^2 \Bigg[ \ga \vel A' \ver^2 + \vel B' \ver^2
\dr \ga 1- x+ 2 r \dr \nnb \\
&&+ \ga \vel C \ver^2 + \vel D \ver^2 \dr \ga 1- x - 4 r \dr \Bigg] 
- 2 C_{10}(1/R) f_B r \int_\delta^{1-4 r} x^2 \,dx \, {\rm Re} \ga A' \dr \,
{\rm ln} 
\displaystyle{\frac{1 + \sqrt{1-\displaystyle{\frac{4 r}{1 - x}}}}
{1 -  \sqrt{1-\displaystyle{\frac{4 r}{1 - x}}}}}
\nnb \\
&&- 4 \vel f_B~ C_{10}(1/R) \ver^2 r \, \frac{1}{m_B^2} \, \int_\delta^{1-4 r} dx
\Bigg[ \ga 2 + \frac{4 r }{x} -\frac{2}{x} -x \dr\,
{\rm ln}
\displaystyle{\frac{1 + \sqrt{1-\displaystyle{\frac{4 r}{1 - x}}}}
{1 -  \sqrt{1-\displaystyle{\frac{4 r}{1 - x}}}}}
\nnb \\
&&+ \frac{2}{x} \ga 1-x \dr \, \sqrt{1-\frac{4 r}{1 - x}}\, \Bigg] \Bigg{\}}~, 
\eea
where the $f_B$ is the leptonic decay constant of the $B$ meson, $x=\displaystyle{\frac{2 E_\gamma}{m_B}}$ is a dimensionless parameter with $E_\gamma$ being the photon energy and
$r=\displaystyle{\frac{m_\tau^2}{m_B^2}}$. The lower limit of integration over $x$ comes from imposing  a cut
on the photon energy (for details see \cite{Aliev2}). Considering  the experimental cut imposed
on the minimum energy for detectable photon, we demand the energy of the
photon to be larger than $50~MeV$, i.e., $E_\gamma \geq a\, m_B$ with
$a \geq 0.01$. As a result, the lower limit is set as $\delta=2 a$ and we will take $\delta=0.02$ for the  lower limit of integration over $x$.
In Eq. (\ref{kabciv}), we have introduced the following coefficients:
\bea
A' =A'(\hat{s},1/R)&=& \frac{A(\hat{s},1/R)}{m_B^2} ~, \nnb \\
B'=B'(\hat{s},1/R)&=& \frac{B(\hat{s},1/R)}{m_B^2} ~,\nnb \\
C=C(1/R)&=& \frac{C_{10}(1/R)}{m_B^2} \,g(q^2)~,\nnb \\
D= D(1/R)&=& \frac{C_{10}(1/R)}{m_B^2} \,f(q^2)~.
\eea

At the end of  this section we would like to present various $1/R$-dependent double--lepton polarization
asymmetries for the transition under study. Note that using the most general model-independent form of the effective Hamiltonian including
all possible forms of the interactions, these double--lepton polarization asymmetries  are calculated in the \cite{veli}. To calculate the double--polarization
asymmetries in our case,  we consider the  polarizations of both lepton and anti-lepton, 
simultaneously and suggest the following  spin projection
operators for the lepton $\ell^-$ and the anti-lepton $\ell^+$: \bea
\Lambda_1 ~\es~ \frac{1}{2} (1+\gamma_5\!\!\not\!{s}_i^-)~,\nnb \\
\Lambda_2 ~\es~ \frac{1}{2} (1+\gamma_5\!\!\not\!{s}_i^+)~,  \eea
 where $i=L,N$ and $T$
correspond to the longitudinal, normal and transversal
polarizations, respectively. Then, we   introduce the following  orthogonal
vectors $s^\mu$ in the rest frame of the lepton and anti-lepton: \bea \label{e6010} s^{-\mu}_L
~\es~ \ga 0,\vec{e}_L^{\,-}\dr =
\ga 0,\frac{\vec{p}_1}{\vel\vec{p}_1 \ver}\dr~, \nnb \\
s^{-\mu}_N ~\es~ \ga 0,\vec{e}_N^{\,-}\dr = \ga
0,\frac{\vec{k}\times
\vec{p}_1}{\vel \vec{k}\times \vec{p}_1 \ver}\dr~, \nnb \\
s^{-\mu}_T ~\es~ \ga 0,\vec{e}_T^{\,-}\dr = \ga 0,\vec{e}_N^{\,-}
\times \vec{e}_L^{\,-} \dr~, \nnb \\
s^{+\mu}_L ~\es~ \ga 0,\vec{e}_L^{\,+}\dr =
\ga 0,\frac{\vec{p}_2}{\vel\vec{p}_2 \ver}\dr~, \nnb \\
s^{+\mu}_N ~\es~ \ga 0,\vec{e}_N^{\,+}\dr = \ga
0,\frac{\vec{k}\times
\vec{p}_2}{\vel \vec{k}\times \vec{p}_2 \ver}\dr~, \nnb \\
s^{+\mu}_T ~\es~ \ga 0,\vec{e}_T^{\,+}\dr = \ga 0,\vec{e}_N^{\,+}
\times \vec{e}_L^{\,+}\dr~, \eea where $\vec{p}_{1(2)}$  are the three--momenta of the leptons $\ell^{-(+)}$ and
$\vec{k}$ is three-momentum of the final photon in the center of mass  (CM) frame  of $\ell^- \,\ell^+$. The longitudinal unit vectors are boosted to the CM frame of $\ell^-
\ell^+$ via Lorenz transformations \bea \label{e6011} \ga s^{-\mu}_L
\dr_{CM} ~\es ~\ga \frac{\vel\vec{p}_1 \ver}{m_\ell}~,
\frac{E \vec{p}_1}{m_\ell \vel\vec{p}_1 \ver}\dr~,\nnb \\
\ga s^{+\mu}_L \dr_{CM} ~\es ~\ga \frac{\vel\vec{p}_1
\ver}{m_\ell}~, -\frac{E \vec{p}_1}{m_\ell \vel\vec{p}_1
\ver}\dr~, \eea while the other two vectors remain unchanged. Finally, we  define the double--lepton polarization asymmetries as: \bea \label{e6012} P_{ij}(\hat{s}) =
\frac{\ds{\Bigg(
\frac{d\Gamma}{d\hat{s}}(\vec{s}_i^-,\vec{s}_j^+)}-
\ds{\frac{d\Gamma}{d\hat{s}}(-\vec{s}_i^-,\vec{s}_j^+) \Bigg)} -
\ds{\Bigg( \frac{d\Gamma}{d\hat{s}}(\vec{s}_i^-,-\vec{s}_j^+)} -
\ds{\frac{d\Gamma}{d\hat{s}}(-\vec{s}_i^-,-\vec{s}_j^+)\Bigg)}}
{\ds{\Bigg( \frac{d\Gamma}{d\hat{s}}(\vec{s}_i^-,\vec{s}_j^+)} +
\ds{\frac{d\Gamma}{d\hat{s}}(-\vec{s}_i^-,\vec{s}_j^+) \Bigg)} +
\ds{\Bigg( \frac{d\Gamma}{d\hat{s}}(\vec{s}_i^-,-\vec{s}_j^+)} +
\ds{\frac{d\Gamma}{d\hat{s}}(-\vec{s}_i^-,-\vec{s}_j^+)\Bigg)}}~,
\eea where the subindex $j$ also stands for the $L,~N$ or $T$ polarization. The  subindexses, $i$ and $j$ correspond
to the lepton and anti-lepton, respectively.
 Using the above  definitions, the various $1/R$-dependent double lepton polarization asymmetries are obtained as:

\bea \label{e6816} P_{LL} (\hat{s},1/R)\es
%0
\frac{1}{\Delta(\hat{s},1/R)} \nnb \\&\times&\Bigg\{ \frac{1}{2} f_B^2 m_B^4
\Big\{ (1-\hat{s})^2 ({\cal I}_1+{\cal I}_4) - [2 \hat{s} +
(1+\hat{s}^2) v^2] {\cal I}_3 + [2 \hat{s} - (1+\hat{s}^2) v^2]
{\cal I}_6 \Big\}
\vel F \ver^2 \nnb \\
&-& \frac{1}{2 \hat{m}_\ell} f_B m_B \hat{s} \Big[ 8 (1+\hat{s})
v^2 + m_B^2 (1-\hat{s}) (2-2 \hat{s} -2 v^2 + 2 \hat{s} v^2 +
v^4 +\hat{s} v^4) {\cal I}_{8} \nnb \\
\ek m_B^2 (1-\hat{s}^2) v^2 {\cal I}_{9}] \Big] \mbox{\rm Re}
[(A_1^\ast + B_1^\ast) F] - \frac{1}{3 \hat{m}_\ell^2} m_B^2
\hat{s}^2 (1-\hat{s})^2 (1-v^2)^2
\mbox{\rm Re} [A_1^\ast B_1 + A_2^\ast B_2] \nnb \\
%17
\ek \frac{2}{3} m_B^2 \hat{s} (1-\hat{s})^2 (1+3 v^2) \Big( \vel
A_1 \ver^2 + \vel A_2 \ver^2 + \vel B_1 \ver^2 + \vel B_2 \ver^2
\Big) \Bigg\}~, \\ \nnb 
\eea
%%%%%%%%%%%%%%%%
\bea
\label{e6817} P_{LN} (\hat{s},1/R)\es
%0
\frac{1}{\Delta(\hat{s},1/R)} \Bigg\{
 f_B m_B^3 \sqrt{\hat{s}} (1-\hat{s}^2) v^2
\mbox{\rm Im} [A_1^\ast F - B_1^\ast F] {\cal I}_7 \nnb \\ &-& 4
\pi f_B m_B \sqrt{\hat{s}} (1-\hat{s}) (1-\sqrt{1-v^2}) \mbox{\rm
Im} [(A_2^\ast + B_2^\ast) F]
\Bigg\}~, \\ \nnb 
\eea
%%%%%%%%%%%%%%%%%%
\bea
\label{e6818} P_{NL} (\hat{s},1/R)\es
\frac{1}{\Delta(\hat{s},1/R)} \Bigg\{ f_B m_B^3 \sqrt{\hat{s}}
(1-\hat{s}^2) v^2
\mbox{\rm Im} [-A_1^\ast F + B_1^\ast F] {\cal I}_7 \nnb \\
&+& 4 \pi f_B m_B \sqrt{\hat{s}} (1-\hat{s}) (1-\sqrt{1-v^2})
\mbox{\rm Im} [- (A_2^\ast + B_2^\ast) F]
\Bigg\}~, \\ \nnb 
\eea
%%%%%%%%%%%%%%%%%%
\bea
\label{e6819} P_{LT}(\hat{s},1/R) \es
\frac{1}{\Delta(\hat{s},1/R)} \Bigg\{ -\frac{1}{\sqrt{\hat{s}}}
f_B^2 m_B^4 \hat{m}_\ell (1-\hat{s}) v \Big[(1+\hat{s}) \vel F
\ver^2 \Big]
({\cal I}_2+{\cal I}_4) \nnb \\
&+& \frac{4}{v} \pi f_B m_B \sqrt{\hat{s}} (1-\hat{s})
(1-\sqrt{1-v^2})
\mbox{\rm Re}[ (A_2^\ast - B_2^\ast) F] + 2 m_B \hat{m}_\ell \mbox{\rm Re}[A_1^\ast A_2 - B_1^\ast B_2] \Big] \nnb \\
&-& \frac{4}{v} \pi f_B m_B \sqrt{\hat{s}} (1+\hat{s})
(1-\sqrt{1-v^2}) \mbox{\rm Re}[(A_1^\ast + B_1^\ast) F] \Big]
\Bigg\}~, \\ \nnb 
\eea
%%%%%%%%%%%%%%%%%
\bea
\label{e6820} P_{TL} (\hat{s},1/R) \es
%0
\frac{1}{\Delta(\hat{s},1/R)} \Bigg\{
%1
- \frac{1}{\sqrt{\hat{s}}} f_B^2 m_B^4 \hat{m}_\ell (1-\hat{s}) v
\Big[(1+\hat{s}) \vel F \ver^2 \Big]
({\cal I}_2+{\cal I}_4) \nnb \\
&-&\frac{4}{v} \pi f_B m_B \sqrt{\hat{s}} (1-\hat{s})
(1-\sqrt{1-v^2}) \mbox{\rm Re}[ (A_2^\ast - B_2^\ast) F] - 2 m_B
\hat{m}_\ell \mbox{\rm Re}[A_1^\ast A_2 - B_1^\ast B_2] \Big] \nnb \\
&-&\frac{4}{v} \pi f_B m_B \sqrt{\hat{s}} (1+\hat{s})
(1-\sqrt{1-v^2})\mbox{\rm Re}[(A_1^\ast + B_1^\ast) F] \Big]
\Bigg\}~, \\ \nnb 
\eea
%%%%%%%%%%%%%%%%
\bea
\label{e6821} P_{NT}(\hat{s},1/R)  \es
%0
\frac{1}{\Delta(\hat{s},1/R)} \Bigg\{ 2 f_B m_B^3 \hat{m}_\ell
(1-\hat{s})^2 v
\mbox{\rm Im}[ -A_1^\ast F + B_1^\ast F]({\cal I}_{8}-{\cal I}_{9}) \nnb \\
&-& 2 f_B m_B^3 \hat{m}_\ell (1-\hat{s}^2) v \mbox{\rm
Im}[(A_2^\ast+B_2^\ast) F] ({\cal I}_{8}-{\cal I}_{9}) \nnb
\\ &-& \frac{8}{3} m_B (1-\hat{s})^2 v \mbox{\rm Im}[- m_B \hat{s}
(A_1^\ast B_1 + A_2^\ast B_2)] \Bigg\}~, \\ \nnb 
\eea
%%%%%%%%%%%%%%%%
\bea
\label{e6822} P_{TN}(\hat{s},1/R)\es
%0
\frac{1}{\Delta(\hat{s},1/R)} \Bigg\{ 2 f_B m_B^3 \hat{m}_\ell
(1-\hat{s})^2 v \mbox{\rm Im}[ A_1^\ast F - B_1^\ast F]
({\cal I}_{8}-{\cal I}_{9}) \nnb \\
&-& 2 f_B m_B^3 \hat{m}_\ell (1-\hat{s}^2) v
\mbox{\rm Im}[(A_2^\ast+B_2^\ast) F] ({\cal I}_{8}-{\cal I}_{9}) \nnb \\
&+&\frac{8}{3} m_B (1-\hat{s})^2 v \mbox{\rm Im}[- m_B \hat{s}
(A_1^\ast B_1 + A_2^\ast B_2)]\Bigg\}~, \\ \nnb 
\eea
%%%%%%%%%%%%%%%%
\bea
\label{e6823} P_{NN} (\hat{s},1/R) \es
%0
\frac{1}{\Delta(\hat{s},1/R)} \Bigg\{ f_B^2 m_B^4 \hat{s} \Big[
(1+v^2) {\cal I}_3 - (1-v^2) {\cal I}_6 \Big]
\vel F \ver^2 \nnb \\
&+& \frac{4}{3} m_B^2 \hat{s} (1-\hat{s})^2 v^2 \Big( 2 \mbox{\rm
Re}[A_1^\ast B_1 + A_2^\ast B_2 ] \Big)\Bigg\}~, \\ \nnb 
\eea
%%%%%%%%%%%%%%%%
\bea
\label{e6824} P_{TT}(\hat{s},1/R)  \es
%0
\frac{1}{\Delta(\hat{s},1/R)} \Bigg\{ \frac{1}{2} f_B^2 m_B^4
\Big\{ - (1-\hat{s})^2 (1-v^2) {\cal I}_1
+ [1 -v^2 - 4 \hat{s} + \hat{s}^2 (1-v^2)] {\cal I}_3 \nnb \\
\ek (1-v^2) (1-\hat{s})^2 {\cal I}_4 + (1-v^2) (1-\hat{s}^2) {\cal
I}_6 \Big\}
\vel F \ver^2 \nnb \\
%5
\ek 4 f_B m_B^3 \hat{m}_\ell (1-\hat{s})^2
\mbox{\rm Re}[(A_1^\ast + B_1^\ast) F] ({\cal I}_{8}-{\cal I}_{9}) \nnb \\
&+& m_B \hat{m}_\ell \Big( \vel A_1 \ver^2 + \vel A_2 \ver^2 +\vel
B_1 \ver^2 +\vel B_1 \ver^2 \Big) \Big] + \frac{8}{3} m_B^2
(1-\hat{s})^2 \Big( \hat{s} \mbox{\rm Re}[A_1^\ast B_1 + A_2^\ast
B_2] \Big) \Bigg\}~,  \nnb \\
\eea
%%%%%%%%%%%%%%%%%%%
where,
\bea
\Delta (\hat{s},1/R) \es \label{e6825}
%1
16 m_B \hat{m}_\ell (1-\hat{s})^2 \Big(\mbox{\rm Re}[m_B
\hat{m}_\ell (A_1^\ast B_1 + A_2^\ast B_1)] \Big) \nnb \\ &+&
\frac{2}{3} (1-\hat{s})^2 \Big[m_B^2 \hat{s} (3+v^2) \Big( \vel
A_1 \ver^2 + \vel A_2 \ver^2 + \vel B_1 \ver^2 + \vel B_2 \ver^2 \Big) \Big] \nnb \\
&-& \frac{1}{2} f_B^2 m_B^4 \vel F \ver^2 \Big\{ (1-\hat{s})^2 v^2
({\cal I}_1 + {\cal I}_4) - (1+\hat{s}^2 + 2 \hat{s} v^2) {\cal
I}_3 - [1-\hat{s} (4 - \hat{s} -2 v^2)] {\cal I}_6 \Big\} \nnb \\
&+& 2 f_B m_B \hat{m}_\ell \mbox{\rm Re} [(A_1^\ast + B_1^\ast) F]
\Big[ 8 (1+\hat{s}) + m_B^2 (1-\hat{s}^2) v^2 {\cal I}_{8} + m_B^2
(1-\hat{s}) (1-3\hat{s}) {\cal I}_{9} \Big] ~. \nnb \\ \eea

In Eqs. (\ref{e6816})--(\ref{e6825}),
$v=\sqrt{1-4 \hat{m}_\ell^2/\hat{s}}$ is the lepton velocity and
\bea \label{e6810}
A_1 = A_1 (\hat{s},1/R) &=& \frac{-2 C_{7}^{eff}(1/R)}{q^2} \Big(m _{b} + m _{d} \Big) g_1(q^2) +  (C_{9}^{eff}(\hat{s},1/R)-C_{10}(1/R)) g(q^2) ~, \nnb \\
A_2 = A_2 (\hat{s},1/R) &=& \frac{-2 C_{7}^{eff}(1/R)}{q^2} \Big(m _{b} - m _{d} \Big) f_1(q^2) +  (C_{9}^{eff}(\hat{s},1/R)-C_{10}(1/R)) f(q^2) ~, \nnb \\
B_1 = B_1 (\hat{s},1/R) &=& \frac{-2 C_{7}^{eff}(1/R)}{q^2} \Big(m _{b} + m _{d} \Big) g_1(q^2) +  (C_{9}^{eff}(\hat{s},1/R)+C_{10}(1/R)) g(q^2) ~, \nnb \\
B_2 = B_2 (\hat{s},1/R) &=& \frac{-2 C_{7}^{eff}(1/R)}{q^2}
\Big(m_{b}-m_{d} \Big) f_1(q^2) + (C_{9}^{eff}(\hat{s},1/R)+C_{10}(1/R)) f(q^2) ~, \nnb\\
 F =F(1/R)&=& 4 m_\ell C_{10}(1/R).\eea
 In the above equations, the ${\cal I}_i$ have the
following representations: \bea {\cal I}_i = \int_{-1}^{+1} {\cal F}_i(z)
dz~,\nnb \eea where \bea
\begin{array}{lll}
{\cal F}_{1}  = \ds\frac{z^2}{(p_1 \cdot k) (p_2\cdot
k)}~,& {\cal F}_{2} = \ds\frac{z}{(p_1 \cdot k) (p_2
\cdot k)}~,&
{\cal F}_{3}  = \ds\frac{1}{(p_1 \cdot k) (p_2 \cdot k)}~, \\ \\
{\cal F}_{4}  = \ds\frac{z^2}{(p_1 \cdot k)^2}~,&
{\cal F}_{5}  = \ds\frac{z}{(p_1 \cdot k)^2}~,&
{\cal F}_{6}  = \ds\frac{1}{(p_1 \cdot k)^2}~, \\ \\
{\cal F}_{7}  = \ds\frac{z}{(p_2 \cdot k)^2}~,& {\cal
F}_{8} = \ds\frac{z^2}{p_1 \cdot k}~,& {\cal
F}_{9} = \ds\frac{1}{p_1 \cdot k}~.
\end{array} 
\eea

\section{Numerical results and discussion}

In this section, we numerically analyze the  physical observables related to the  radiative dileptonic $B \rar \gamma l^+ l^- $ transition both in the ACD and  SM models. The main input parameters entering
the calculations are form factors. These form factors  which we will use in our numerical calculations have been calculated using the
light cone QCD sum rules \cite{Aliev1,Eilam2,Alievsav,Buchalla}:
\bea
\begin{array}{ll}
g_(q^2) = \ds\frac{1~GeV}{\ds \ga
1-\frac{q^2}{(5.6~GeV)^2}\dr^2}~,&
f_(q^2) = \ds\frac{0.8~GeV}{\ds \ga 1-\frac{q^2}{(6.5~GeV)^2}\dr^2}~,\\ \\
g_1(q^2) = \ds\frac{3.74~GeV^2}{\ds \ga
1-\frac{q^2}{40.5~GeV^2}\dr^2}~,& f_1(q^2) =
\ds\frac{0.68~GeV^2}{\ds \ga 1-\frac{q^2}{30~GeV^2}\dr^2}~.
\end{array} \nnb
\eea

We also use some other input values to numerically analyze the branching ratios as well as various double---lepton polarization asymmetries: $m_{t}=167~GeV$, $m_{W}=80.4~GeV$, $m_{Z}=91.18~GeV$,
$m_{c}=1.46~GeV$, $m_{b}=4.8~GeV$, $m_{u}=0.005~GeV$,
$m_{B}=5.28~GeV$, $\alpha_{em}=\frac{1}{137}$,   $| V_{tb}
V_{td}^* |= 0.01$, $G_F = 1.167\times10^{-5}~GeV^{-2}$, $m_e =
5.1\times10^{-4}~GeV$, $m_\mu=0.109~GeV$, $m_\tau=1.784~GeV$, and
$\tau_B=1.525\times10^{-12}s$ .

We start by presenting the results on branching ratios. As we previously noted, in the case of $e$ and $\mu$ as final leptons, we only consider the contribution of the structure dependent part to the amplitude.
Integrating Eq. (\ref{kab}) over $\hat{s}$ in the whole physical region, $4\hat{m_l}^2\leq \hat{s}\leq 1$ we obtain the total $1/R$-dependent decay width for the $l=e$ or 
$\mu$. Multiplying this result by
the lifetime of the $B$ meson in appropriate unit, we acquire the branching ratio as a function of the compactification factor of extra dimension. As the results of the $e$ and $\mu$ channels have similar behavior 
and are close to each other, we only depict the numerical results in the $\mu$ channel. In figure 1, we present the sensitivity of the branching ratio in
$\mu$ channel on the compactification factor both in the UED and SM models  in the interval $200~GeV\leq 1/R\leq 1000~GeV$. 

Few comments about the lower bound of the compactification scale are in order. Analysis of the $B\rightarrow X_s\gamma$ transition  and also anomalous magnetic moment
  had  previously shown  that the experimental data are in a good consistency with the ACD model if $1/R\geq 300~GeV$ \cite{ikiuc}. From the electroweak precision tests, the lower limit for $1/R$ had also previously been  obtained to be $%
250~GeV$  if $M_h\geq250~GeV$ representing larger KK contributions to the low energy FCNC processes,
 and $300~GeV$  if $M_h\leq250~GeV$ \cite{ACD,yee}. 
 Using also the 
electroweak precision measurements as well as
some cosmological constraints, the authors of \cite{Gogoladze:2006br} and \cite{Cembranos:2006gt} have shown that the lower limit on compactification factor is in or above the $500~GeV$. Taking into account the leading order
(LO) contributions due to the exchange of KK modes together with the available next-to-next-to-leading order (NNLO) corrections 
to also $B(B \rightarrow X_s \gamma)$ decay channel in the SM, the authors of \cite{Haisch:2007vb} have found a lower bound on the inverse compactification radius as $600~GeV$.  Finally, 
using final states with jets and missing transverse momentum, the ATLAS collaboration at CERN is set a $600~GeV$ on the lower limit of the compactification scale
for values of the compression scale between $2$ and $40$, translating to a lower bound of $730~ GeV$ on the mass of the KK gluon \cite{ATLAS}.
 We will plot the physical observables
under consideration in the range $1/R\in[200-1000]GeV$ just to clearly show how the results of the UED deviate from those of the SM and grow decreasing the value of $1/R$.
\begin{figure}
\begin{center}
\includegraphics[scale=0.8]{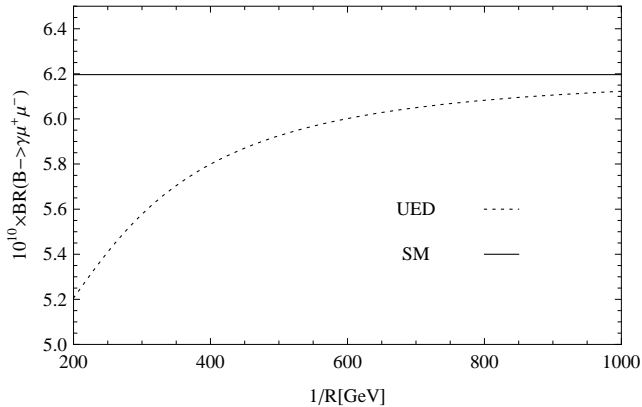}
\end{center}
\caption{ The dependence of the branching ratio for  $B \rar \gamma \mu^+ \mu^- $ on the compactification factor, $1/R$. \label{fig1}}
\end{figure}

From  figure 1 we see that
\begin{itemize}
 \item there is a considerable discrepancy between the UED and the SM predictions at lower values of the compactification factor. When $1/R$ is increased the result 
of UED approaches to that of the SM, such that at 
upper limit of $1/R$, two models have approximately the same predictions. This discrepancy can be considered as a signal for existing  extra dimensions in nature   we should look for in hadron colliders.
\item Although the contribution of single universal extra dimension has different effects on the Wilson coefficients such that the $ C_{10}$ is enhanced and the $ C_7^{eff}$  is suppressed (for details see \cite{nihan}), 
the branching ratio is suppressed at small values of $1/R$. This is against the effect of the UED in some semileptonic decay channels considered for instance in \cite{kank1,nihan}.
\item The order of the branching ratios in both models indicate  that the predicted results lie below the  existing experimental upper limit, $Br(B \rar \gamma \mu^+ \mu^-)<1.6\times10^{-7}$ \cite{Nakamura}. 
\end{itemize}

Now, we proceed to depict the results of both UED and SM models on the branching ratio of the $B \rar \gamma \tau^+ \tau^- $ decay channel. Using the differential decay rate in Eq. (\ref{kabciv}) which
 contains both the structure-dependent and Bremsstrahlung parts, we obtain the results for branching ratio as shown in figure 2. Here  we also see a sizable
difference between the predictions of two models at lower values of the compactification factor. Increasing the value of $1/R$ leads to an increase in the 
value of the branching ratio such that at upper limit, the result
 of UED becomes roughly the same as the SM. The orders of branching ratios show that the  decay channel under consideration is more probable in $\tau$ channel case compared to that of the
 $\mu$. With 
developments at LHC, we hope we will be able to detect these channels and determine the exact branching ratios.
\begin{figure}
\begin{center}
\includegraphics[scale=0.8]{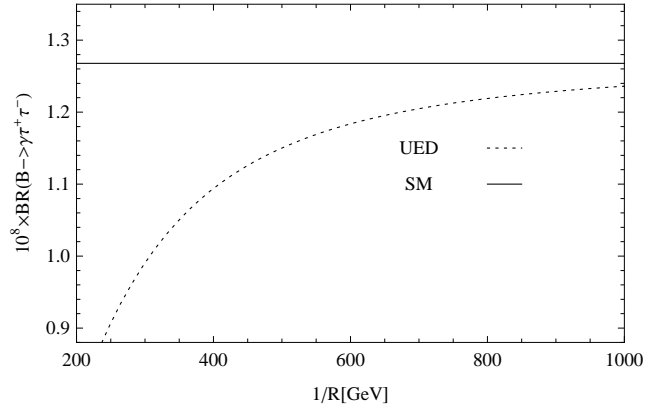}
\end{center}
\caption{ The dependence of the branching ratio for  $B \rar \gamma \tau^+ \tau^- $ on the compactification factor, $1/R$
 \label{fig1p}}
\end{figure}
\begin{figure}
\begin{center}
\includegraphics[scale=0.7]{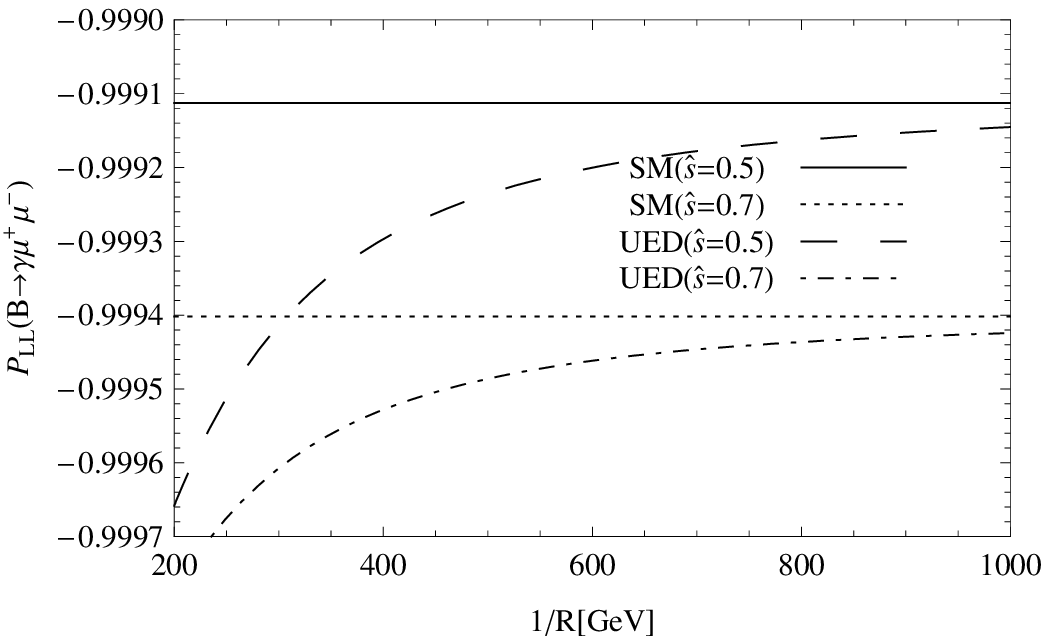}
\includegraphics[scale=0.7]{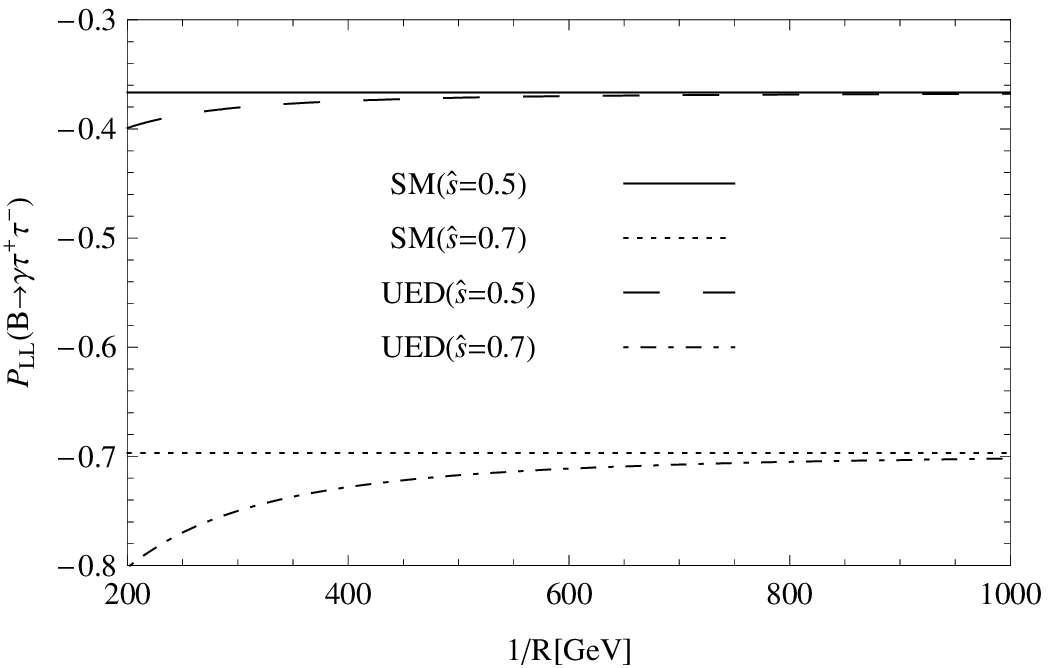}
\end{center}
\caption{ The dependence of the $P_{LL}$ polarization in two models for $B \rar \gamma l^+ l^- $ on the
compactification factor, $1/R$, for both leptons and at two fixed values of
$\hat{s}$. \label{fig2}}
\end{figure}
\begin{figure}
\begin{center}
\includegraphics[scale=0.7]{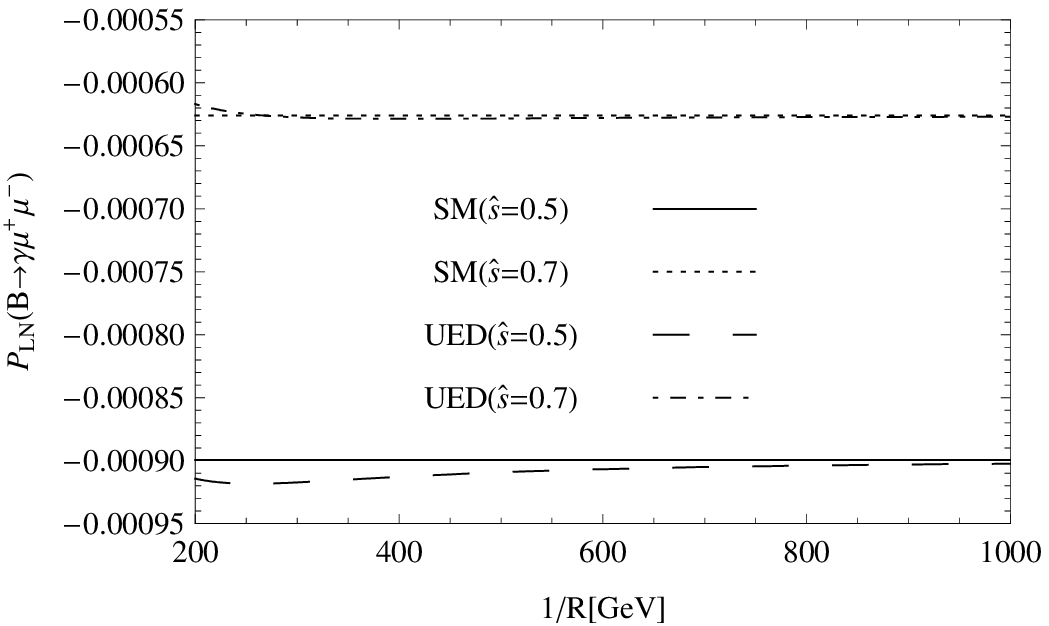}
\includegraphics[scale=0.7]{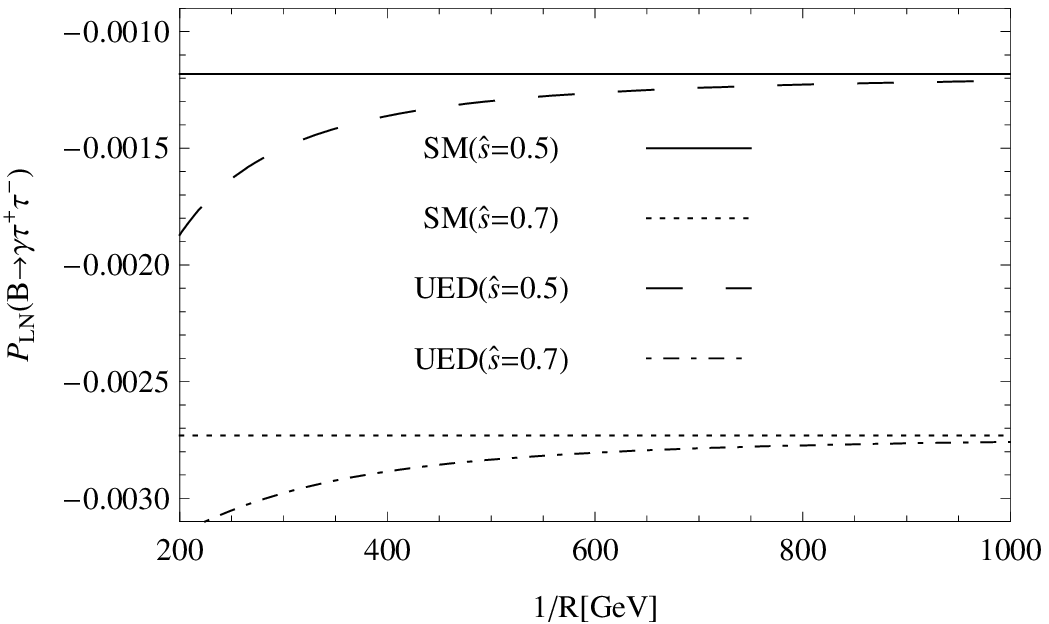}
\end{center}
\caption{ The same as Figure \ref{fig2}, but for $P_{LN}$. \label{fig2p}}
\end{figure}
\begin{figure}
\begin{center}
\includegraphics[scale=0.7]{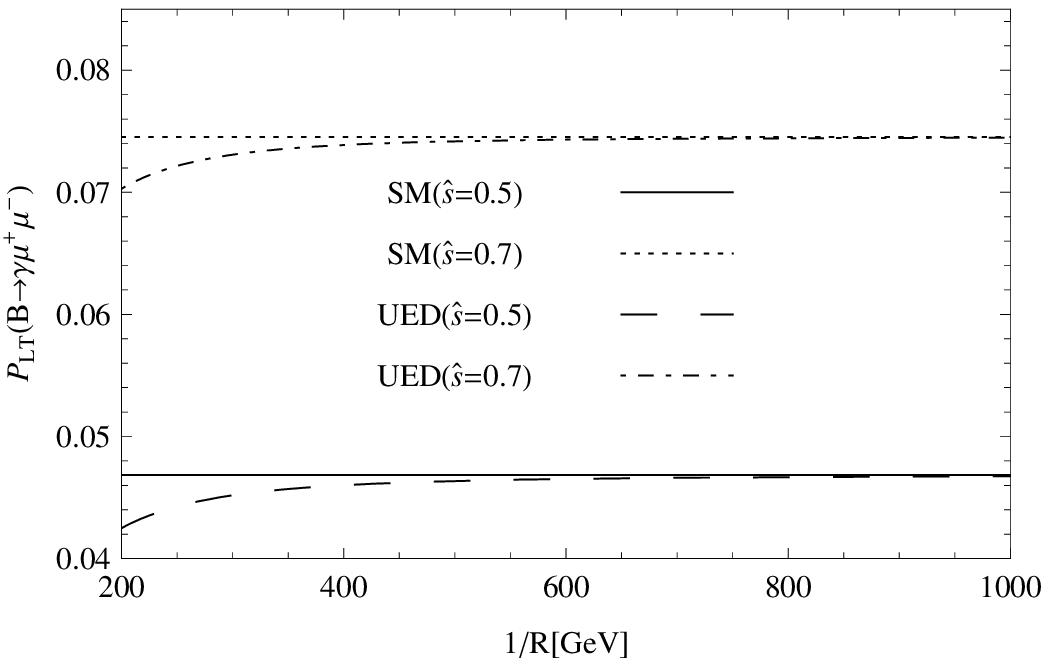}
\includegraphics[scale=0.7]{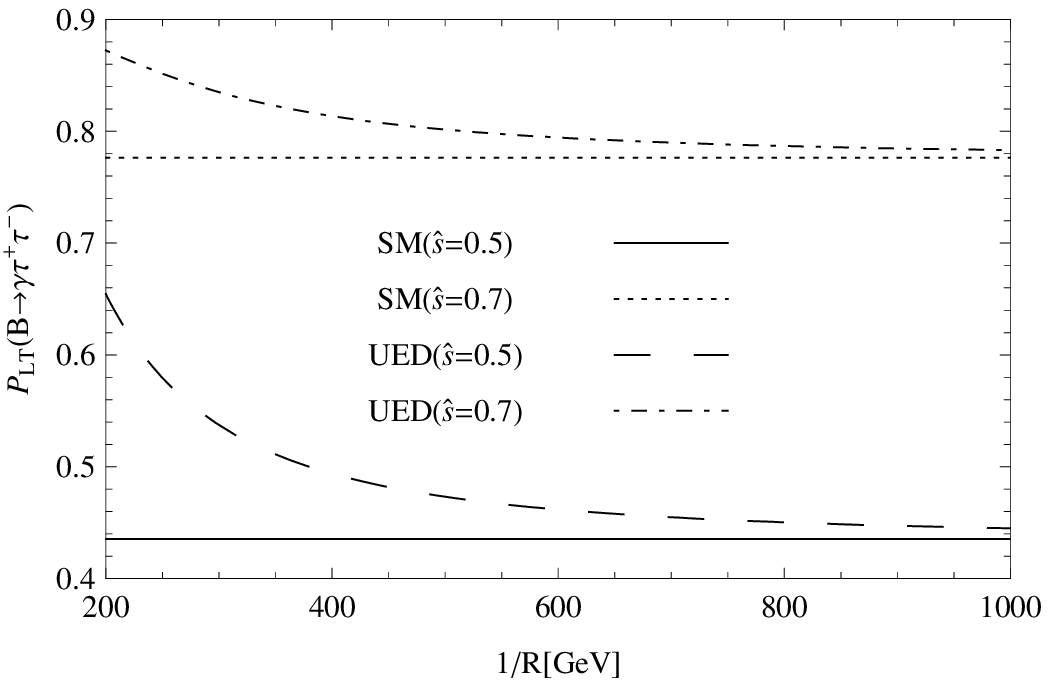}
\end{center}
\caption{ The same as Figure \ref{fig2}, but for $P_{LT}$. \label{fig2pp}}
\end{figure}
\begin{figure}
\begin{center}
\includegraphics[scale=0.7]{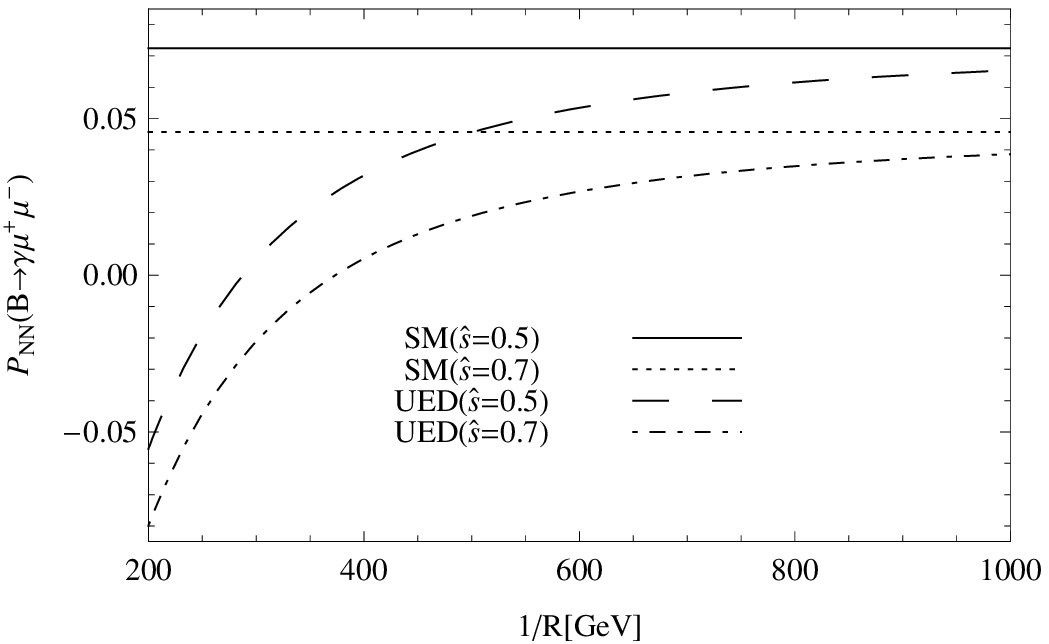}
\includegraphics[scale=0.7]{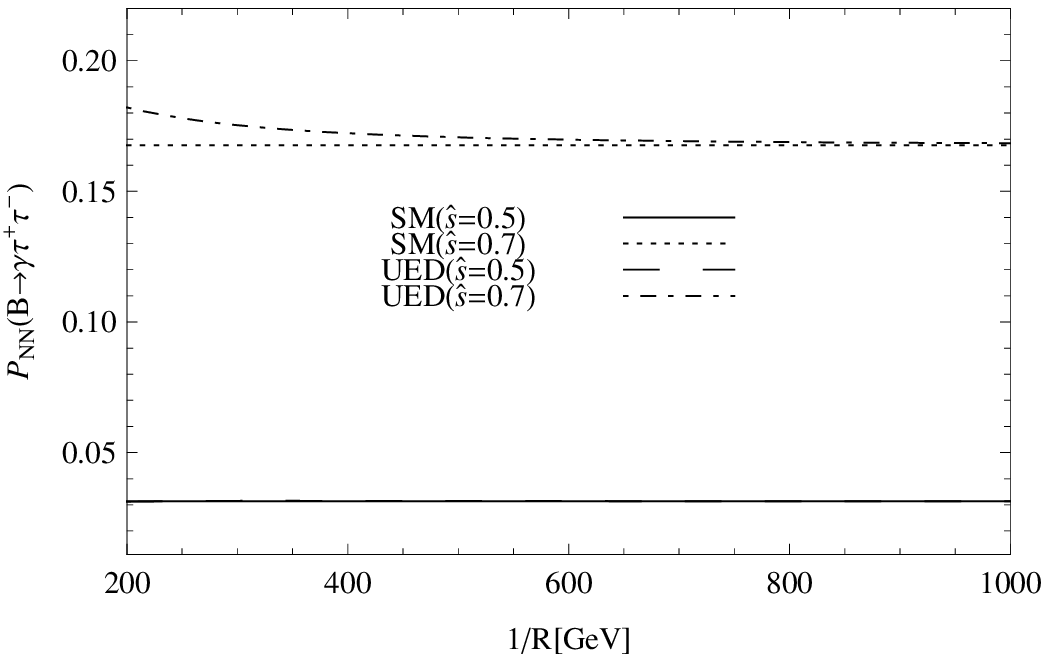}
\end{center}
\caption{ The same as Figure \ref{fig2} but for $P_{NN}$ \label{fig3}}
\end{figure}
\begin{figure}
\begin{center}
\includegraphics[scale=0.7]{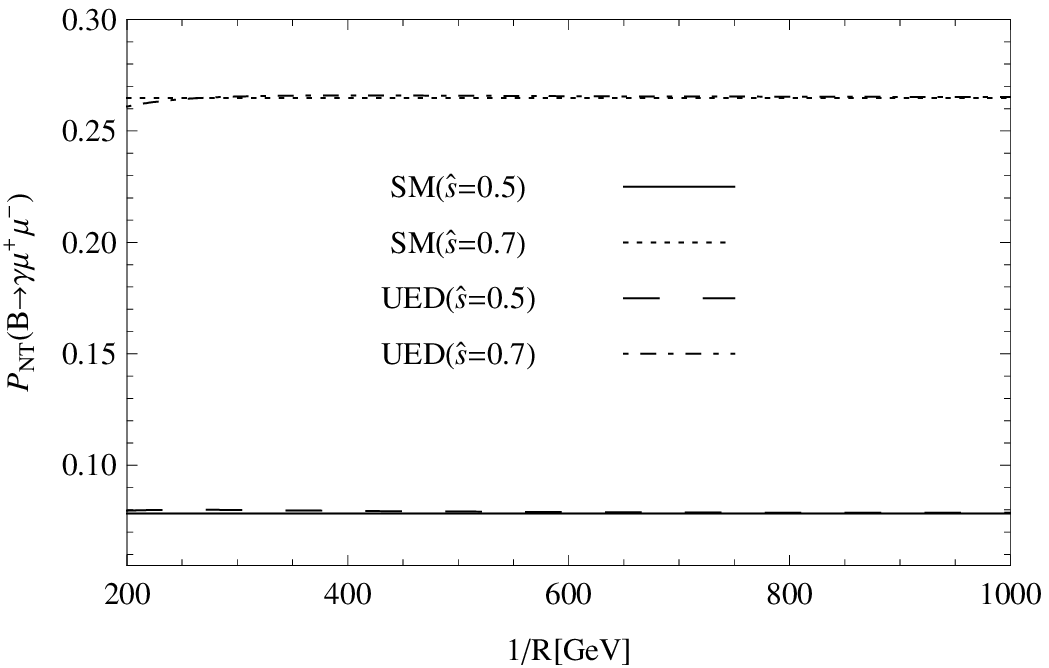}
\includegraphics[scale=0.7]{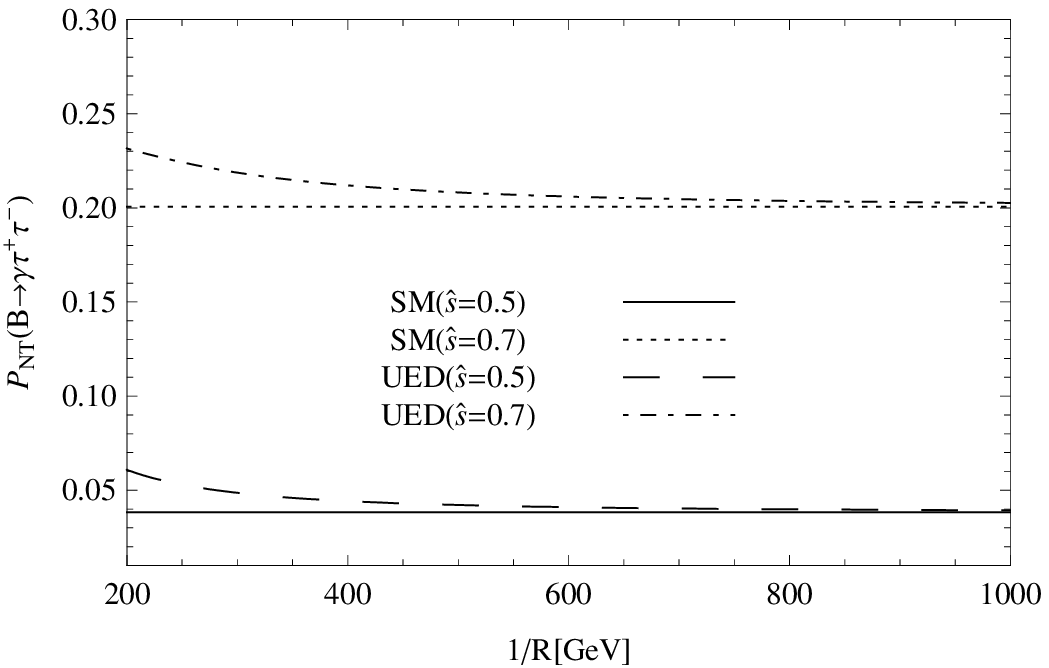}
\end{center}
\caption{ The same as Figure \ref{fig2}, but for $P_{NT}$ \label{fig3p}}
\end{figure}
\begin{figure}
\begin{center}
\includegraphics[scale=0.7]{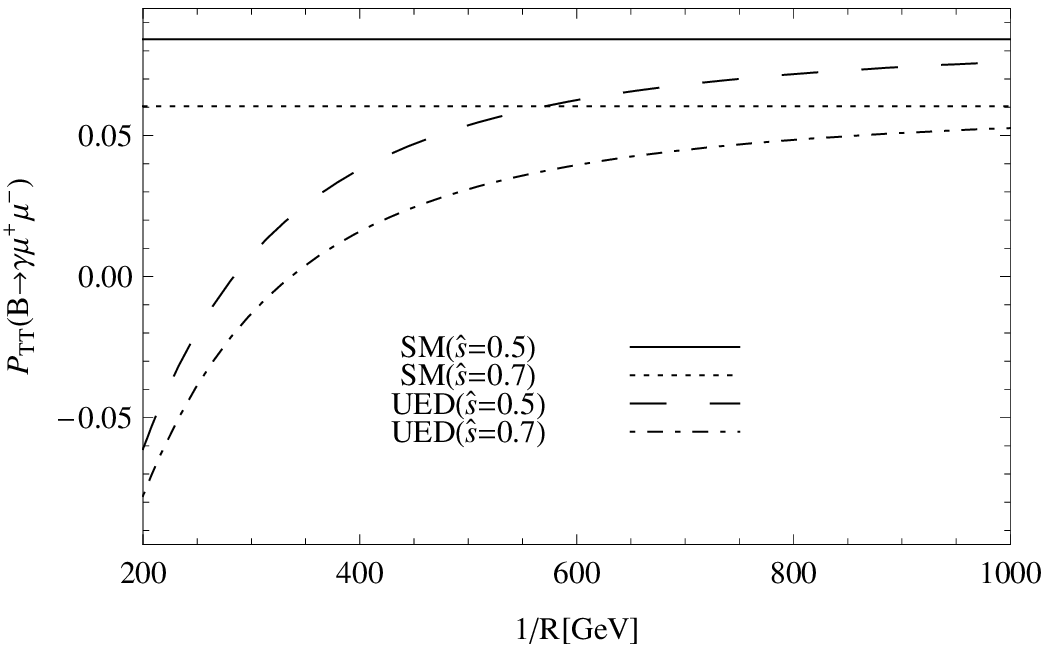}
\includegraphics[scale=0.7]{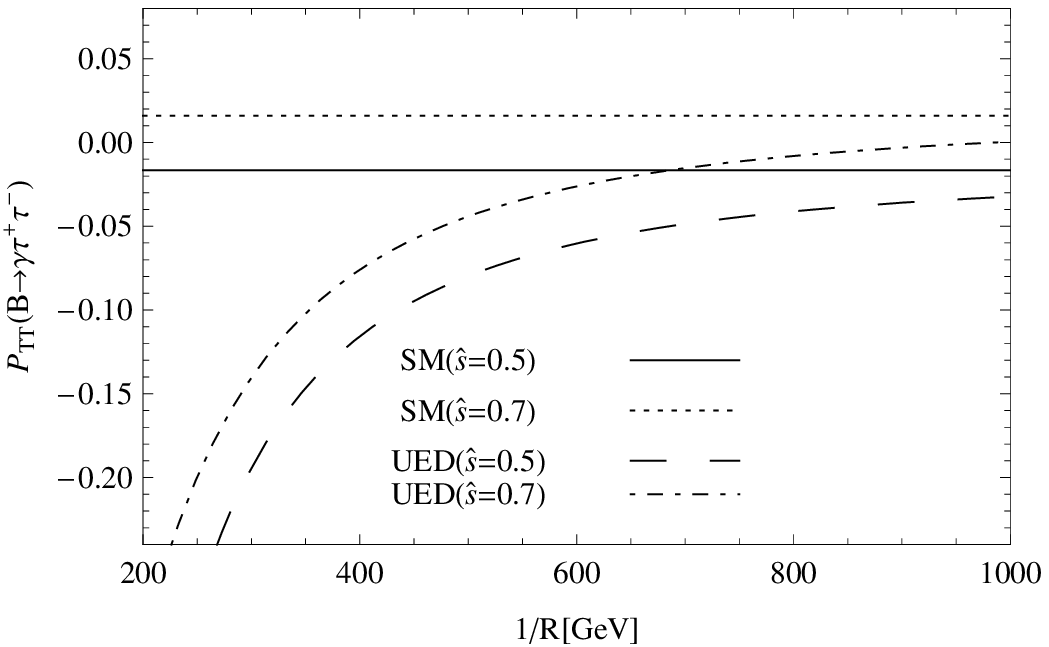}
\end{center}
\caption{ The same as Figure \ref{fig2}, but for $P_{TT}$ \label{fig3pp}}
\end{figure}

At the end of this section, we would like to show our numerical results on the various double-lepton polarization asymmetries considered in the previous section. For $P_{i\neq j}$, we only show one of the 
polarizations, not both $P_{ij}$ and  $P_{ji}$. We depict the sensitivity of the various double-lepton polarization asymmetries associated with the radiative dileptonic $B \rar \gamma l^+ l^- $ decay channel
on the compactification factor of extra dimension  in figures \ref{fig2}-\ref{fig3pp}. As it is clear from the explicit expressions for the double--lepton polarization asymmetries in the previous section, they have
both dependencies on the $1/R$ and the $\hat{s}$. Here, we depict the results for the compactification factor dependence at two fixed  values of the $\hat{s}$ in its allowed region. With a quick glance at these figures,
 we see that
\begin{itemize}
 \item the polarizations $P_{TT}$, $P_{LT}$, $P_{LN}$ and $P_{LL}$ in the $\tau$ channel as well as the $P_{TT}$ and $P_{NN}$ in the $\mu$ channel show considerable discrepancies between predictions of the two models
 at lower values of the compactification factor. For the remaining cases, those differences are small. 
\item Some of the double--lepton polarization asymmetries like $P_{TT}$ in the $\tau$ channel are very small  in the SM; however take  sizable values in the UED model specially at lower values of the compactification factor.
\item At a fixed value of $1/R$, we detect  strong dependencies on  $\hat{s}$ for the  $P_{LL}$, $P_{LN}$, $P_{LT}$, $P_{NN}$ and $P_{NT}$ in the $\tau$ 
channel as well as the $P_{LN}$,  $P_{LT}$ and $P_{NT}$ in the $\mu$ channel.
\end{itemize}
Our concluding remark is that any measurement of the  physical observables considered in the present study and the comparison of these data with our predictions can give valuable information about the nature of existing extra dimensions.

\section{Acknowledgment}
We would like to thank T. M. Aliev for  useful discussions.

\end{document}